\documentclass{article}
\usepackage[dvips]{graphics}
\title{
Quantization of the free charge carriers on InSb
at room temperature
}
\author{
A.M. Yafyasov, I.M. Ivankiv and V.B. Bogevolnov \\[10pt]
\small\em Department of Solid State Electronics, Institute of Physics,\\
\small\em St.Petersburg State University, \\
\small\em Ulyanovskaya 1, 198904 St.Petersburg, Russia\\
\small\em E-mail: yafyasov@pobox.spbu.ru
}
\date{}
\begin{document}
\maketitle

\abstract{

A new method of self-consistent quantum calculation
of the density of the space charge near the surface of a crystal
is carried out for the semiconductor with nonparabolic (Kane)
dispersion law of bands.
The remarkable feature is the solution of the Schr\"odinger
equation for electrons and holes in the energy range, including
both bound energy states and states in the continuum.
Theoretical voltage-capacitance dependence is calculated and coincides
with experimental data. The dependence of the electron mass
and surface mobility from the value of surface potential are analyzed.

}

\section{Method of self-consistent quantum calculation}

When an electric field is applied perpendicularly to the surface
of the semiconductor, the space charge region (SCR) is induced
in the subsurface layer \cite{Ando}.
In frames of the one-particle Hartree approximation the distribution
of charge density is defined from self-consistent solution of Poisson
and Schr\"odiner equations \cite{Stern1}.
In InSb at room temperature it is necessary
to take into account simultaneously both type of charge
carriers in the whole allowed range of energies.
Therefore the calculation is based on the model of
``semi-infinite crystal with a second fictitious border" \cite{pss}
with modifications implied a specific character of Kane semiconductors.

The bulk dispersion law of Kane semiconductors is described by formula \cite{Kane}
\begin{equation}
k^{2}(E)=\frac{1}{P^{2}}
\frac{(E-E_{0})(E-E_{0}+E_{g})(E-E_{0}+E_{g}+\Delta)}{(E-E_{0}+E_{g}+\frac{2}{3}\Delta)},
\label{e_disp}
\end{equation}
where
$\mbox{\bf{k}}$ is wave vector,
$P$ is the matrix element, accouting the interaction between
the conductance and valence bands,
$E_{0}=\hbar^{2}k^{2}/(2m_{0})$ is the energy of free electron with mass $m_{0}$.
$E_{g}$ is the energy gap of semiconductor,
$\Delta$ is the energy of spin-orbital splitting of valence band.

In general case the density of charge carriers in the bulk of semiconductor
is determined by Fermi-Dirac statistic
\begin{equation}
\rho = \frac{2}{(2\pi)^{3}}\int f_{0}({\bf k})\,d{\bf k},
\label{fr}
\end{equation}
where
$$
f_{0}({\bf k})=\frac{1}{1+\exp[(E({\bf k})-E_{F})/(k_{0}T)]}.
$$
$E_{F}$ is the Fermi level energy;
$k_{0}$ is the Boltzmann constant;
$T$ is the absolute temperature.

Consider the crystal with the bounded coordinate
$z\in[0,L^{*}]$ and the infinited coordinates
$x,y\in]-\infty, +\infty[$. In frames of the one-particle Har\-tree
approximation the wave function may be presented
by $e^{i{\bf k}_{\parallel}{\bf r}_{\parallel}}\varphi_{i}(z,k_{\parallel})$,
where $\varphi_{i}(z,k_{\parallel})$ is the enveloping wave function
for the bounded state $E_{i}(k_{\parallel})$.
Here are $k_{\parallel}^{2}=k_{x}^{2}+k_{y}^{2}$,
$k_{x},k_{y}\in]-\infty,+\infty[$; $r_{\parallel}^{2}=x^{2}+y^{2}$.
After the change of variables
$\varepsilon_{\parallel}={\hbar^{2}k_{\parallel}^{2}}/({2m_{n}k_{0}T})$
$\varepsilon_{i}=E_{i}/(k_{0}T)$, $\varepsilon_{F}=E_{F}/(k_{0}T)$,
$\varepsilon_{g}=E_{g}/(k_{0}T)$
the concentrations of the electrons and hard holes
can be found from expressions
\begin{equation}
\rho_{e}(z) =
\frac{m_{n}k_{0}T}{\pi \hbar^{2}}\int\limits_{0}^{\infty}
d\varepsilon_{\parallel}
\sum\limits_{i}\frac{|\varphi_{i}(z,\varepsilon_{\parallel})|^{2}}%
{1+\exp (\varepsilon_{i}-\varepsilon_{F})}
\label{qe}
\end{equation}
\begin{equation}
\rho_{hh}(z) = \sum\limits_{j=1}^{+\infty}
\Gamma_{j}(\varepsilon_{j})|\varphi_{j}(z)|^{2},
\label{qhh}
\end{equation}
where
$$
\Gamma_{j}(\varepsilon_{j})=\frac{m_{hh}k_{0}T}{\pi\hbar^{2}}
\ln\left(1+
\exp(-\varepsilon_{j}-\varepsilon_{g}-\varepsilon_{F})\right).
$$

Electrostatical potential into the SCR $V(z)$
created by external electric field satisfies the Poisson equation \cite{pss}
\begin{equation}
\frac {d^{2} V (z)} {d\, z^{2}} =
q \cdot \frac {\rho_{e} (z) -\rho_{hh}(z) + N_{a} - N_{d}}
{\varepsilon_{0} \varepsilon_{sc}},
\quad V (0) =V_{s};\ V (L^{*}) = V^{0}(L^{*}).
\label{i3}
\end{equation}
Here $\varepsilon_{sc}$ is the static dielectric
constant of the semiconductor; $N_{d}$ and $N_{a}$ are the concentrations
of ionized donor and acceptor impurities, respectively.

In terms of quantum description in the one-particle Hartree
approximation the wave functions
$\varphi_{i}(z,k_{\parallel})$, $\varphi_{j}(z)$
and eigenvalues $E_{i}(k_{\parallel}$, $E_{j}$
of the free charge carriers are derived from solving of the equations
\begin{equation}
-\frac{\partial^{2} \varphi(z,k_{\parallel})}{\partial z^{2}} =
[k^{2}(E-V(z))-k_{\parallel}^{2}]\varphi(z,k_{\parallel})
\label{e_schr}
\end{equation}
\begin{equation}
\left(-\frac {\hbar^{2}} {2 m_{hh}} \frac {d^{2}} {d\, z^{2}} - qV(z)
\right) \varphi_{j} (z) = E_{j} \varphi_{j} (z),
\label{shhh}
\end{equation}
Boundary conditions and normalization of all wave functions take the form
$$
\varphi(0) = \varphi(L^{*}) = 0;\
\int\limits_{0}^{L^{*}} | \varphi(z)|^{2} \, dz=1.
$$
The densities of charge carriers are calculated according to
Eqs.(\ref{qe}) and (\ref{qhh}).
In our case at room temperature the integration is performed
in the interval $\varepsilon_{\parallel}\in [0,16]$ taking into
account all quantum subbands up to $\varepsilon_{i}(k_{\parallel}=0)|<12$
and $|-\varepsilon_{j}-\varepsilon_{g}|<12$ for electrons and holes
respectively. The self-consistent calculation of SCR is carried out
in frames of the iteration scheme \cite{pss}.
\begin{figure}
\includegraphics{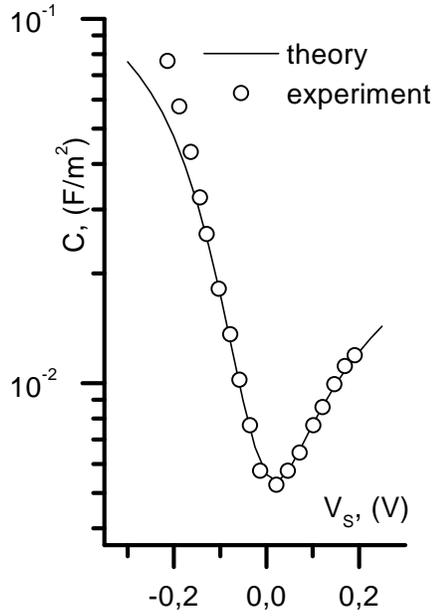}
\caption{Voltage-capacitance dependences.
\label{fig1}}
\end{figure}

\section{Experimental results and discussion}

The following constants of the material \cite{Kik} were used for calculations:
$\varepsilon_{sc}=17.9$; $m_{0}=9.1\cdot 10^{-31}\,$kg;
$m_{n}=0.013m_{0}$; $m_{hh}=0.5m_{0}$;
$N_{a,d}=0$; $E_{g}=0.165-2.8\cdot 10^{-4}(T-300)\,$eV; $T=290\,$K.
The voltage-capacitance dependence is an important measurable
characteristics. The differential capacitance of SCR is given by
$$
C_{sc}(V_{s})=\frac{d\,Q}{dV_{s}},\quad \mbox{where}\ \
Q = q\cdot\int \limits_{0}^{\infty}[\rho_{e}(z)-\rho_{hh}(z) + N_{a} - N_{d}]\, dz.
$$
The comparison between experimental data \cite{Rom} and
theoretical quantum calculation
for InSb at room temperature is shown on Fig.\,\ref{fig1}.
Good agreement of the experiment and quantum calculation
in wide range of surface potentials demonstrates the efficiency
of proposed mathematical model.

\begin{figure}
\includegraphics{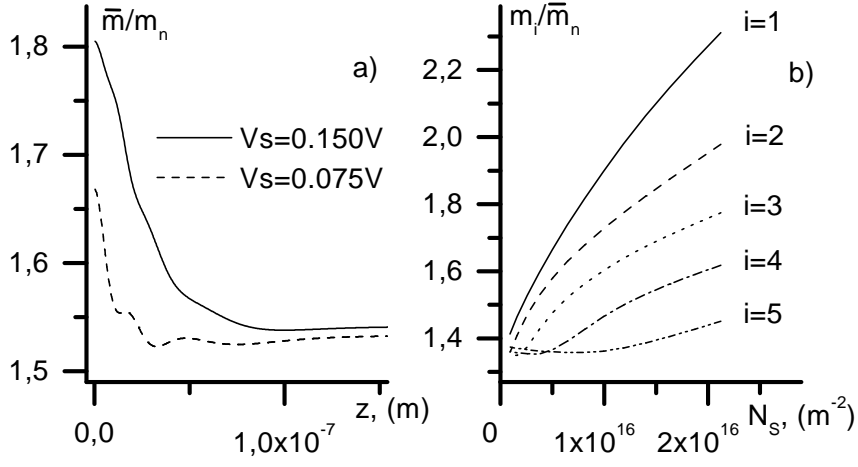}
\caption{
{\em a\/} --- the mean mass of electron on depth;
{\em b\/} --- the mean mass of bounded electron.
\label{fig2}
}
\end{figure}

This model of calculation allows to carry out the detailed analysis of
the change of the electron mass in SCR for the Kane semiconductor.
The mass of electron in longitude direction is described by formula
$$
\frac{1}{m_{i}(E_{\parallel})} =
\frac{1}{\hbar^{2}k_{\parallel}}\frac{dE_{i}}{dk_{\parallel}}=
\frac{1}{m_{n}}\frac{dE_{i}}{dE_{\parallel}}.
$$
Determine the mean value of the electron mass $\overline{m}(z)$ as
$$
\overline{m}(z)=
\left.\int\limits_{0}^{\infty} d\varepsilon_{\parallel}
\sum\limits_{i}\frac{m_{i}(\varepsilon_{\parallel})|\varphi_{i}(z,\varepsilon_{\parallel})|^{2}}{1+\exp (\varepsilon_{i}-\varepsilon_{F})}\right/ \rho_{e}(z)
$$
One can see from Fig.\,\ref{fig2}{\em a\/}
that near the surface the concentration
of the ``hard" electron is greater than in the bulk.

On the Fig.\,\ref{fig2}{\em b\/} is shown the average mass of electrons
on $i$ energy level
$$
\overline{m}_{i} =
\left.\int\limits_{0}^{\infty}\frac{m_{i}(\varepsilon_{\parallel})\,d\varepsilon_{\parallel}}{1+\exp(\varepsilon_{i}-\varepsilon_{F})}\right/
\int\limits_{0}^{\infty}\frac{d\varepsilon_{\parallel}}{1+\exp(\varepsilon_{i}-\varepsilon_{F})}.
$$

Let us introduce the mean mass of electron in SCR as
$$
m_{SCR}(V_{s}) =
\left.\int\limits_{0}^{2L_{D}}
\overline{m}(z)\rho_{e}(z)\,dz\right/\int\limits_{0}^{2L_{D}} \rho_{e}(z)\,dz,
$$
where $L_{D}$ is the length of Debye.
\begin{figure}
\includegraphics{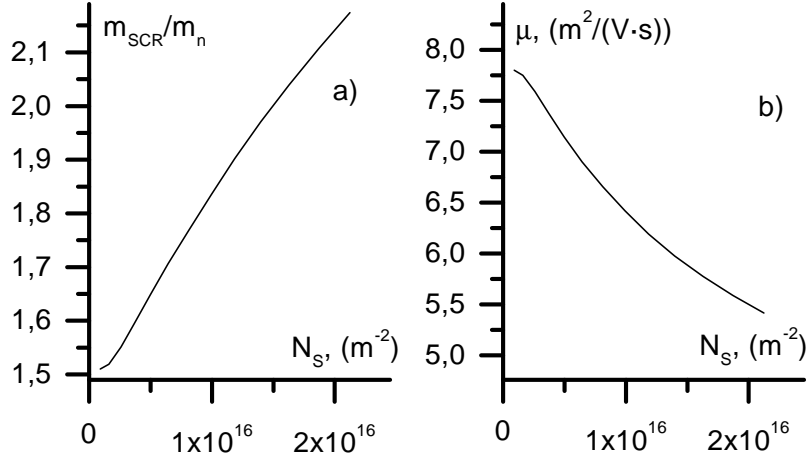}
\caption{
{\em a\/} --- the mean mass of the electron in SCR;
{\em b\/} --- surface mobility of the electron in SCR.
\label{fig3}
}
\end{figure}

This dependence is shown on Fig.\,\ref{fig3}{\em a}.
It is obvious that the mean mass of electron increases monotonically
with growth of surface potential.
This regularity must arise also in transport phenomena
in subsurface layer of Kane semiconductor.

Indeed, let us write the expression of the electron mobility
in the bulk of the semiconductor and in SCR by $m_{SCR}(V_{s})$:
$$
\mu_{b}=q\tau_{b}/m_{SCR}(V_{s}=0),
\qquad
\mu_{s}=q\tau_{s}/m_{SCR}(V_{s}).
$$
Assume that there is no additional mechanisms of scattering on the surface
other than the ones in the bulk ($\tau_{s}=\tau_{b}$). Then we have
$$
\mu_{s}(V_{s}) = \mu_{b}\displaystyle\frac{m_{SCR}(V_{s}=0)}{m_{SCR}(V_{s})}.
$$
The dependence of surface mobility on
$N_{s}=\int\limits_{0}^{L^{*}}
(\rho_{e}(z) -\rho_{hh}(z) + N_{a} - N_{d})\,dz$
is shown on Fig.\,\ref{fig3}{\em b\/},
where $\mu_{b}=7.8$\,m$^{2}$/(V$\cdot$s)
\cite{Kik}. As one can see from Fig.\,\ref{fig3}{\em b\/}, for InSb at room
temperature the surface mobility, even in absence of additional mechanisms
of surface scattering (interaction with phonon, roughnesses, etc \cite{Ando})
decreases monotonically with growth of the surface potential.
This effect is stipulated by nonparabolic dispersion law of conductance band,
which leads to the dependence of electron mass on energy and surface
potential.

The results can be very useful for studying
of mobility of electrons and holes in the accumulation
and inversion layers on the surface of the Kane semiconductors.

{\bf Acknowledgements.}
We thank to Commission of European Communities for financial support
in frames of EC--Russia Exploratory Collaborative Activity under
Contract ESPRIT Project NTCONG 28890.


\begin{thebibliography}{99}
\bibitem {Ando} T. Ando, A.B. Fowler, F. Stern, Rev. Mod. Phys.
                438 (1982) 54.
\bibitem {Stern1} F. Stern, W.E. Howard, Phys. Rev.  816 (1967) 163.
\bibitem{pss} A.M. Yafyasov, I.M. Ivankiv, phys. stat. sol.(b)
                 (1998) 208.
\bibitem{Kane} O.E. Kane, J. Phys. Chem. Sol.   249 (1957) .
\bibitem{Kik} I.S. Grigoriev, Z.P. Melikhov (Eds.),
 Physical sizes, Energoatomizdat, Moscow (in Russian) 1991, p.750.
\bibitem{Rom} O.V. Romanov, A.M. Yafyasov,
                phys. stat. sol.(a)  235 (1986) 94.
\end{thebibliography}
\end{document}